\RequirePackage[2020-02-02]{latexrelease}
\documentclass[aip,graphicx]{revtex4-1}
\usepackage{graphicx}
\usepackage{epsfig}
\usepackage{color}
\usepackage{epstopdf}
\usepackage{amsmath}
\usepackage{caption}
\usepackage{subcaption}

\usepackage[normalem]{ulem}

\newcommand{\blue}{\textcolor{black}}

\begin{document}


\title{Suppression of reconnection in polarized, thin magnetotail current sheets: 2D simulations and implications}



\author{Xin An}
 \thanks{Author to whom correspondence should be addressed}
 \email{xinan@epss.ucla.edu}
\affiliation{
Department of Earth, Space and Planetary Sciences, University of California, Los Angeles, CA, 90095, USA.
}%

\author{Anton Artemyev}
\affiliation{
Department of Earth, Space and Planetary Sciences, University of California, Los Angeles, CA, 90095, USA.
}%
\affiliation{Space Research Institute of the Russian Academy of Sciences, Moscow, 117997, Russia}

\author{Vassilis Angelopoulos}
\affiliation{
Department of Earth, Space and Planetary Sciences, University of California, Los Angeles, CA, 90095, USA.
}%

\author{Andrei Runov}
\affiliation{
Department of Earth, Space and Planetary Sciences, University of California, Los Angeles, CA, 90095, USA.
}%

\author{San Lu}
\affiliation{
School of Earth and Space Sciences, University of Science and Technology of China, Hefei, 230026, China.
}%

\author{Philip Pritchett}
\affiliation{
Department of Physics and Astronomy, University of California, Los Angeles, CA, 90095, USA.
}%

\date{\today}

\begin{abstract}
Many in-situ spacecraft observations have demonstrated that magnetic reconnection in the Earth's magnetotail is largely controlled by the pre-reconnection current sheet configuration. One of the most important thin current sheet characteristics is the preponderance of electron currents driven by strong polarized electric fields, which are commonly observed in the Earth's magnetotail well before the reconnection. We use particle-in-cell simulations to investigate magnetic reconnection in the 2D magnetotail current sheet with a finite magnetic field component normal to the current sheet and with the current sheet polarization. Under the same external driving conditions, reconnection in a polarized current sheet is shown to occur at a lower rate than in a nonpolarized current sheet. The reconnection rate in a polarized current sheet decreases linearly as the electron current's contribution to the cross-tail current increases. In simulations with lower background temperature the reconnection electric field is higher. We demonstrate that after reconnection in such a polarized current sheet, the outflow energy flux is mostly in the form of ion enthalpy flux, followed by electron enthalpy flux, Poynting flux, ion kinetic energy flux and electron kinetic energy flux. These findings are consistent with spacecraft observations. Because current sheet polarization is not uniform along the magnetotail, our results suggest that it may slow down reconnection in the most polarized near-Earth magnetotail and thereby move the location of reconnection onset downtail.
\end{abstract}


\maketitle

\section{Introduction\label{sec:introduction}}
Substorm onset, an unsolved problem in magnetospheric physics \cite{Baker96,Angelopoulos08,Lu20:natcom}, can be reformulated as concerning magnetic field-line reconnection in the Earth's magnetotail current sheet \cite{Sitnov19}. One of the important questions is: what are the factors determining where (at what distance from the Earth) magnetic reconnections occur? Statistical spacecraft observations of current sheet dynamics \cite{Nagai13:statistics,Genestreti14,Lu20:natcom} show magnetic reconnection predominantly in the middle magnetotail ($\sim 20-35\,\mathrm{Re}$ downtail, $\,\mathrm{Re}$ being the Earth radius), whereas the reconnection occurs much rarely in the near-Earth ($<15\,\mathrm{Re}$) magnetotail \cite{Sergeev08,Angelopoulos20}. Theory and simulations should reveal factors controlling the location of reconnection onset.

\begin{figure*}[tphb]
    \centering
    \includegraphics[width=\textwidth]{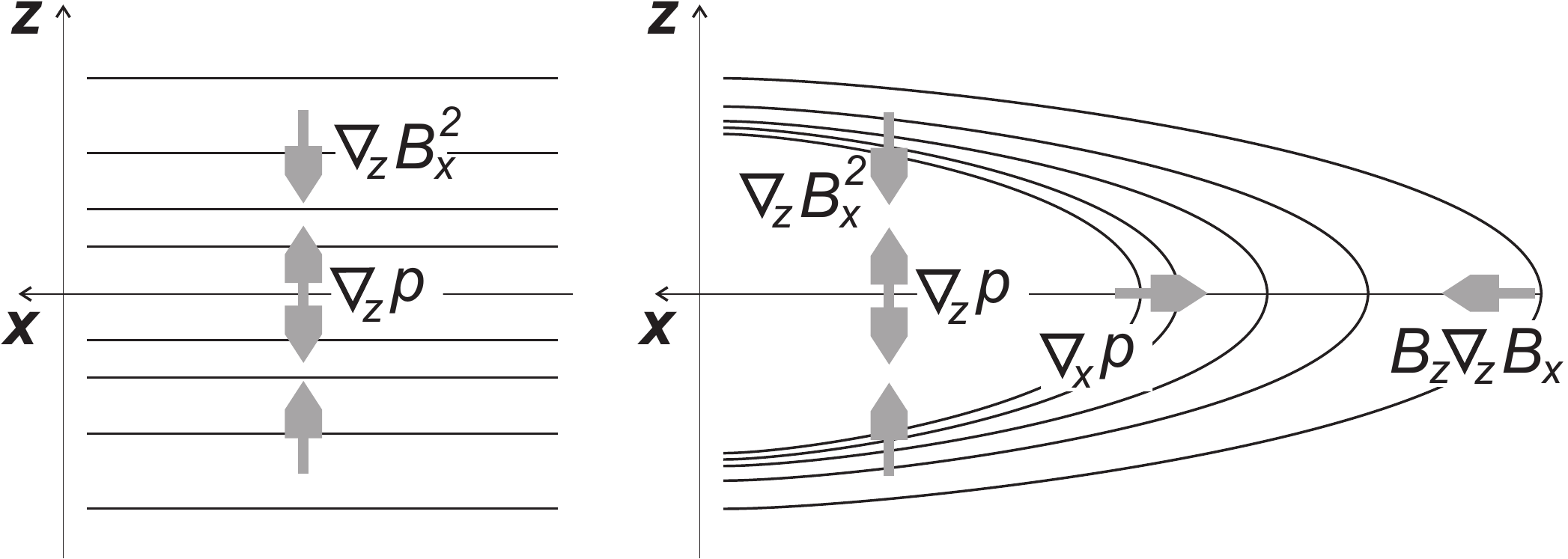}
    \caption{Schematic view of 1D tangential discontinuity (left) and 2D magnetotail current sheet (right). Black curves show magnetic field lines. Grey arrows show principal components of the stress balance: plasma pressure gradient $\nabla p$ and magnetic field tension force $\sim ({\bf B} \cdot \nabla {\bf B})$.}
    \label{fig:schematic}
\end{figure*}

Properties of the magnetotail current sheet strongly vary with the radial distance: current density, plasma density, ion and electron temperatures, and magnetic field magnitude are higher closer to the Earth \cite{Artemyev16:jgr:pressure,Artemyev17:jgr:THEMIS&Geotail,Birn21:AGU,Runov21:jastp}. However, reconnection onset is rarely observed in the most intense (with a larger current density), near-Earth current sheets, but mostly seen in the middle tail \cite{Nagai13:statistics,Genestreti14}. In contrast to the magnetopause current sheets often described as one-dimensional (1D) tangential discontinuities \cite{Roth96,Mozer&Pritchett11,Allanson17:grl}, the magnetotail current sheet is essentially two-dimensional (2D) with a finite normal magnetic field (see schematic Figure \ref{fig:schematic}). This normal magnetic field produces a magnetic field tension force balanced by plasma pressure gradient along the magnetotail \cite{Schindler72,LP82}, which differs the magnetotail current sheet from the simpler 1D tangential discontinuities \cite{Harris62}. The finite normal magnetic field component stabilizes the current sheet configuration relative to the tearing instability and prevents the spontaneous reconnection \cite{Pellat91,Quest96,Sitnov10}. Thus, the reconnection onset in such 2D current sheet is determined by the interplay between the external driver and the 2D current sheet configuration (see discussion in Refs.~\onlinecite{Sitnov02,Zelenyi08JASTP,Sitnov&Merkin16}). Such external driver forces the magnetotail current sheet thinning (so-called substrom growth phase \cite{Baker96,Angelopoulos08}). This process determines the pre-reconnection current sheet properties \cite{Lu18:3dthinning,Sitnov21:grl,An22:currentsheet} and may control the reconnection onset location. Therefore, accurate simulation of the pre-reconnection thinning is important for the investigation of magnetotail current sheet reconnection. 

The external driver thins the magnetotail current sheet prior the reconnection, and such thinning takes about one hour \cite{Petrukovich07,Sun17:cs_pressure}. At the end of the current sheet thinning, right before the reconnection onset, spacecraft detect thin ion-scale current sheet \cite{Artemyev16:jgr:thinning}. The configuration of this ion-scale pre-reconnection sheet likely determines the properties of magnetotail reconnection. Although the current sheet thinning can be reproduced in hybrid \cite{Lu16:cs,Runov21:jastp} and magnetohydrodynamic (MHD) \cite{Birn98:cs,Merkin15, Nishimura&Lyons16:flows} simulations, this process is too long to be modeled using particle-in-cell (PIC) simulations, which typically start with already sufficiently thin (ion-scale) pre-reconnection current sheets \cite{Pritchett01, Sitnov11,Sitnov13, Liu14:CS, Pritchett&Lu18}. Models of such ion-scale current sheets contain strong diamagnetic ion currents carried by ion flow with a speed comparable to the ion thermal speed \cite{Schindler72,LP82}. However, such strong ion flows are absent in realistic magnetotail current sheet, which has been explained due to the current sheet polarization during thin current sheet formation in Ref.~\onlinecite{Hesse98}. Current sheet thinning requires a current density increase, which implies an ion-to-electron cross-field drift increase, but ion acceleration is energy consuming. Such drift is easily produced by the motion of lighter electrons caused by an ${\bf E}\times{\bf B}$ drift, which does not produce new currents (although it can, see Refs.~\onlinecite{Lu16:cs}), but compensates ion gradient drifts (making ions nearly nonmoving in the laboratory reference frame) and increases electron drifts \cite{Pritchett&Coroniti95,Zelenyi10GRL}. The electric field of polarized current sheet are the same Hall electric field that forms in the reconnection region due to separation of ion and electron motions, but in the magnetotail current sheet such field should exist well before the reconnection onset.

Cluster \cite{Escoubet01} and THEMIS \cite{Angelopoulos08:ssr} missions have confirmed the absence of strong ion currents in thin ion-scale current sheets \cite{Runov05,Artemyev09:angeo}, whereas accurate measurements of electron currents and 3D electric field by MMS mission \cite{Burch16,Torbert18} have demonstrated the presence of both strong electron currents \cite{Lu19:jgr:cs,Richard21} and polarization electric fields \cite{Wang18:mms,Hubbert21,Leonenko21Ge&Ae}. Note that, during the pre-MMS era, direct measurements of electron currents and sub-ion scale current sheets were almost impossible (see discussion in Ref.~\onlinecite{Nakamura06}), and main statistical results of current sheet polarization (detected as an absence of ion currents in thin current sheet with strong total current density estimated from magnetic field gradients \cite{Dunlop02,Runov05:pss}) were obtained for ion-scale pre-reconnection current sheets \cite{Runov05,Artemyev09:angeo}. MMS observations in the Earth's magnetotail confirmed the results of ion-scale current sheet polarization \cite{Lu19:jgr:cs,Richard21}, but also revealed a new class of sub-ion scale (or electron-scale) current sheets \cite{Wang18:mms, Hubbert21}. Therefore, there are formally two classes of thin current sheets in the magnetotail: ion-scale polarized current sheets (almost no ion currents, but strong contribution of ion thermal pressure to the stress balance and strong electron currents) and electron-scale current sheets (without any ion contribution to the current sheet structure). The relation of these two classes is yet unknown: electron-scale current sheets may result from further thinning of ion-scale current sheets right before the reconnection onset \cite{Lu20:natcom,Lu22:grl} or can be {\it relic} of the reconnection \cite{Nakamura21,Leonenko21}.  In this paper we focus on the impact of ion-scale current sheet configuration on the reconnection properties, whereas formation, structure, and stability of electron-scale current sheets require further theoretical and observational studies.

The current sheet polarization and the absence of ion currents are intrinsic properties of thin current sheets, and shall be included into initial current sheet configuration in PIC simulations. Spacecraft observations show that such polarization electric field is essentially localized around the center of current sheets \cite{Wang18:mms, Hubbert21,Richard21}. Theoretical models explain this localization as an effect of a cold background plasma observed at the current sheet boundaries \cite{Haaland08, Andre15, Toledo-Redondo21} and reducing the polarization field ${\bf E}$ to zero there \cite{YL04,Birn04, Schindler12}. Thus, the ${\bf E}\times{\bf B}$ drift is not a simple transformation of the reference frame for magnetotail current sheets, as it would be for a constant ${\bf E}\times{\bf B}$ drift speed. Ref.~\onlinecite{Lu20:pop} has shown that such nonuniform ${\bf E}\times{\bf B}$ drift changes the current sheet configuration and may affect reconnection in 1D current sheet model \cite{Harris62, YL04}, whereas Refs.~\onlinecite{Panov&Pritchett18:rippling, Sitnov21:grl} have investigated post-reconnection dynamics of initially polarized 2D current sheets. In this study, we generalize the investigation of the role of current sheet polarization in Ref.~\onlinecite{Lu20:pop} to 2D magnetotail current sheets. 

If the current sheet polarization affects the reconnection onset in the magnetotail, this effect may control the onset location, because the polarization intensity varies with the radial distance \cite{Artemyev16:jgr:ex}. Such variation is caused by differences between the ion-to-electron temperature ratio in the distant tail ($>60\,\mathrm{Re}$ covered by ARTEMIS and Geotail orbits; where electrons are $10$ times colder than ions) and that ratio in the near-Earth tail ($7$-$15\,\mathrm{Re}$ covered by THEMIS orbits after 2010; where electrons are only $2-3$ times colder than ions) \cite{Wang12,Runov18}. Thus, although the ion-to-electron ${\bf E}\times{\bf B}$ drift imbalance can supply most of the current in the distant tail, ion and electron diamagnetic drifts contribute significantly to the current in the near-Earth tail. In thin current sheets, comparable to or less than the thermal ion gyroradius, ions are particularly inefficient in supplying the needed current density, and electron currents (due to ${\bf E}\times{\bf B}$ drift imbalance, polarization, or electron diamagnetism) provide the main contribution to the spatial variation and dynamics of the current density \cite{Artemyev09:angeo,Lu19:jgr:cs,Richard21,Hubbert21}. 

In this study we aim to analyze if the polarization of the initial (ion-scale) current sheet may affect the reconnection sufficiently important to determine the location of the reconnection onset in the Earth's magnetotail. For this reason we perform a parametric investigation of the magnetic reconnection in polarized thin current sheets with magnetic field profiles representative of the Earth's magnetotail. We modify a traditional 2D current sheet model \cite{Schindler72,LP82} by including polarization fields that are self-consistent with the development of a reduced ion current density and an increased electron current density (see Section \ref{sec:model}). Although this model does not describe all observed magnetotail current sheet configurations (see discussion in Refs.~\onlinecite{Sitnov&Merkin16, Artemyev21:grl}), it does contain their main features such as a finite normal magnetic field and isotropic plasma pressure (see discussion in Refs.~\onlinecite{Sitnov11,Sitnov21}). We compare reconnection in 2D polarized (with different polarization strength) and nonpolarized (with dominant ion diamagnetic current) thin current sheets (see Section \ref{sec:simulation}). We conclude with a discussion of our results in the context of the Earth's magnetotail (see Section \ref{sec:discussion}).

\section{Current sheet model\label{sec:model}} 
We generalize the equilibrium current sheet of Lembege and Pellat \cite{LP82} to capture the strong, thin electron currents observed in the Earth's magnetotail. In the following, the $x$-axis is along the Sun-Earth line, the $y$-axis is parallel to the dawn-dusk direction (positive duskward) and the $z$-axis completes the right-handed coordinate system. We assume the background magnetic field is in the $x - z$ plane and write it as $\mathbf{B} = \nabla \times A_y(x, z) \mathbf{e}_y$, i.e.,
\begin{equation}
	B_x = -\frac{\partial A_y}{\partial z}, \\
	B_z = \frac{\partial A_y}{\partial x}, \\
	B_y = 0 ,
\end{equation}
where $A_y \mathbf{e}_y$ is the vector potential. Our description also allows for an electrostatic potential $\varphi (x, z)$ that describes static electric fields in the $x$-$z$ plane. The Hamiltonian of a particle of species $\alpha$ moving in this electromagnetic field is
\begin{equation}
	H_\alpha = \frac{p_x^2 + p_z^2}{2 m_\alpha} + \frac{1}{2 m_\alpha} \left(p_y - q_\alpha \frac{A_y}{c} \right)^2 + q_\alpha \varphi .
\end{equation}
Here $\mathbf{p} = m_\alpha \mathbf{v} + q_\alpha \mathbf{A} / c$ is the canonical momentum. Because the Hamiltonian does not explicitly depend on position $y$ and time $t$, the particle motion has two invariants, $p_y$ and $H_\alpha$.

Based on the invariants $p_y$ and $H_\alpha$, we consider the so-called Harris distribution \cite{Harris62}, namely, a drifting Maxwellian with inhomogeneous density
\begin{equation}
	f_\alpha (p_y, H_\alpha) = n_\alpha C \exp\left( - \frac{H_\alpha - v_{D \alpha} p_y}{T_\alpha} \right) ,
\end{equation}
where $n_\alpha$ is the particle density at ($A_y=0$, $\varphi=0$), $v_{D \alpha}$ is the drift velocity, and $T_\alpha$ is the temperature. $C$ is the normalization constant that ensures
\begin{equation}
	\int f_\alpha \mathrm{d}^3v \bigg|_{A_y = 0,\, \varphi = 0} = n_\alpha.
\end{equation}
Calculating the charge and current densities from the zeroth- and first-order moments of $f_\alpha$, respectively, and putting them into Maxwell's equations, we can determine $\varphi$ and $A_y$ self-consistently as
\begin{eqnarray}
	\Delta \varphi = - 4 \pi \sum_{\alpha} q_\alpha n_\alpha \exp\left( -\frac{q_\alpha \varphi}{T_\alpha} + \frac{v_{D\alpha} q_\alpha A_y}{c T_\alpha} \right) , \label{varphi-poisson-boltzmann}\\
	\Delta A_y = - 4 \pi \sum_{\alpha} q_\alpha n_\alpha \frac{v_{D \alpha}}{c} \exp\left( -\frac{q_\alpha \varphi}{T_\alpha} + \frac{v_{D\alpha} q_\alpha A_y}{c T_\alpha} \right) , \label{Ay-poisson-boltzmann}
\end{eqnarray}
where $\Delta = \partial^2 / \partial x^2 + \partial^2 / \partial z^2$ is the Laplace operator in two dimensions. The magnetotail current sheet thickness is much greater than the Debye length, and thus Equation \eqref{varphi-poisson-boltzmann}, Gauss's law, can be replaced by the quasi-neutrality condition, as is common in the low-frequency, long-wavelength approximation, i.e., when the speeds of the wave modes of interest are much lower than the speed of light:
\begin{equation}
	\sum_{\alpha} q_\alpha n_\alpha \exp\left( -\frac{q_\alpha \varphi}{T_\alpha} + \frac{v_{D\alpha} q_\alpha A_y}{c T_\alpha} \right) = 0 \label{eq-quasi-neutrality} .
\end{equation}
To verify the validity of the latter approximation, we ensure that
\begin{equation}
	\vert \Delta \varphi \vert \ll \vert q_\alpha n_\alpha \vert \exp\left( -\frac{q_\alpha \varphi}{T_\alpha} + \frac{v_{D\alpha} q_\alpha A_y}{c T_\alpha} \right)
\end{equation}
a posteriori. In an electron-ion plasma with $q_i = - q_e = e$, one can choose a coordinate system in which
\begin{equation}\label{eq:neutrality-condition}
	\frac{v_{Di}}{T_i} = -\frac{v_{De}}{T_e} ,
\end{equation}
so the current sheet is nonpolarized, i.e., $\varphi = 0$, as is done in the derivation of the Harris-sheet approximation. In the magnetotail, where $T_i > T_e$, Equation \eqref{eq:neutrality-condition} implies that the ion drift velocity is greater than the electron drift velocity, i.e., $\vert v_{Di} / v_{De} \vert > 1$. However, as discussed in Section \ref{sec:introduction}, this condition does not hold for the magnetotail thin current sheet, which is typically polarized. In fact, consistent with observations of a bipolar electric field in the $z$ direction, evidence of the polarization effect, the electron drift velocity in the magnetotail is larger than that of ions \cite{Runov06,Artemyev09:angeo,Lu19:jgr:cs}, i.e., $\vert v_{Di} / v_{De} \vert < 1$. Moreover, in the magnetotail, Equation \eqref{eq:neutrality-condition} cannot be satisfied by choosing the reference frame \cite{YL04,Sitnov21:grl}.

We assume hereafter that each species has two components: a current sheet component (i.e., the current-carrying one) and a background component (i.e., the non-current-carrying one). This background component represents the non-current-carrying plasma population in the lobe of magnetotail. The number density and current density of species $\alpha$, respectively, are specified as
\begin{eqnarray}
	n_\alpha &=&  n_0 \exp\left( -\frac{q_\alpha \varphi}{T_{\alpha 0}} + \frac{v_{D\alpha} q_\alpha A_y}{c T_{\alpha 0}} \right) + n_b \exp\left( -\frac{q_\alpha \varphi}{T_{\alpha b}} \right) , \\
	j_{y \alpha} &=& q_\alpha n_0 v_{D \alpha} \exp\left( -\frac{q_\alpha \varphi}{T_{\alpha 0}} + \frac{v_{D\alpha} q_\alpha A_y}{c T_{\alpha 0}} \right) ,
\end{eqnarray}
where the subscripts ``0'' and ``b'' denote the current sheet component and the background component, respectively. Using these parameters, we define the asymptotic magnetic field $B_0$ at $z \to \pm\infty$ through the pressure balance equation
\begin{equation}
	\frac{B_0^2}{8 \pi} = n_0 (T_{e0} + T_{i0}) .
\end{equation}

The Earth's magnetotail current sheet is characterized by rather stretched magnetic field lines with $\partial^2/\partial x^2 \ll \partial^2/\partial z^2$ (see, e.g., Ref.~\onlinecite{Artemyev21:grl}). Thus, to model the configuration of the Earth's magnetotail, we assume that $A_y$ has a weak dependence on $x$ (see, e.g., Ref.~\onlinecite{Schindler72, LP82}). To be more precise, $A_y$ has the form $A_y = A_y (\varepsilon x, z)$ with $\varepsilon \ll 1$. With this approximation, Equation \eqref{Ay-poisson-boltzmann} may be rewritten as
\begin{equation}
	\frac{\partial^2 A_y}{\partial z^2}  = - 4 \pi \sum_{\alpha} q_\alpha n_0 \frac{v_{D \alpha}}{c} \exp\left( -\frac{q_\alpha \varphi}{T_{\alpha 0}} + \frac{v_{D\alpha} q_\alpha A_y}{c T_{\alpha 0}} \right) \label{eq-Ay-Lembege-Pellat},
\end{equation}
where $\partial^2 A_y / \partial x^2$ is omitted and the equation is exact to order $\varepsilon$. The electrostatic potential $\varphi$ is determined by the quasi-neutrality condition
\begin{equation}\label{eq-quasi-neutrality-2}
		\sum_{\alpha} q_\alpha n_0 \exp\left( -\frac{q_\alpha \varphi}{T_{\alpha 0}} + \frac{v_{D \alpha} q_\alpha A_y}{c T_{\alpha 0}} \right) + q_\alpha n_b \exp\left( -\frac{q_\alpha \varphi}{T_{\alpha b}} \right) = 0.
\end{equation}
The two boundary conditions are specified at $z=0$ as $B_x = 0$ and $B_z=\varepsilon B_0$:
\begin{equation}\label{eq-boundary-condition}
	\begin{split}
		\frac{\partial A_y}{\partial z} \bigg\vert_{z = 0} = 0 &,\\ A_y \big\vert_{z = 0} = \varepsilon B_0 x &.
	\end{split}
\end{equation}
By solving Equations \eqref{eq-Ay-Lembege-Pellat}, \eqref{eq-quasi-neutrality-2} and \eqref{eq-boundary-condition}, we can obtain $A_y$ and $\varphi$ in the rectangular domain $[-L_z/2 \leq z \leq L_z/2]$ and $[-L_x \leq x \leq 0]$, from which the density distributions can be further obtained. Note that we only need to solve for the upper half plane $z > 0$. The solution for the lower half plane $z < 0$ can be obtained by the symmetry $A_y (-z) = A_y (z)$ and $\varphi (-z) = \varphi (z)$.

\begin{figure*}[tphb]
    \centering
    \includegraphics[width=\textwidth]{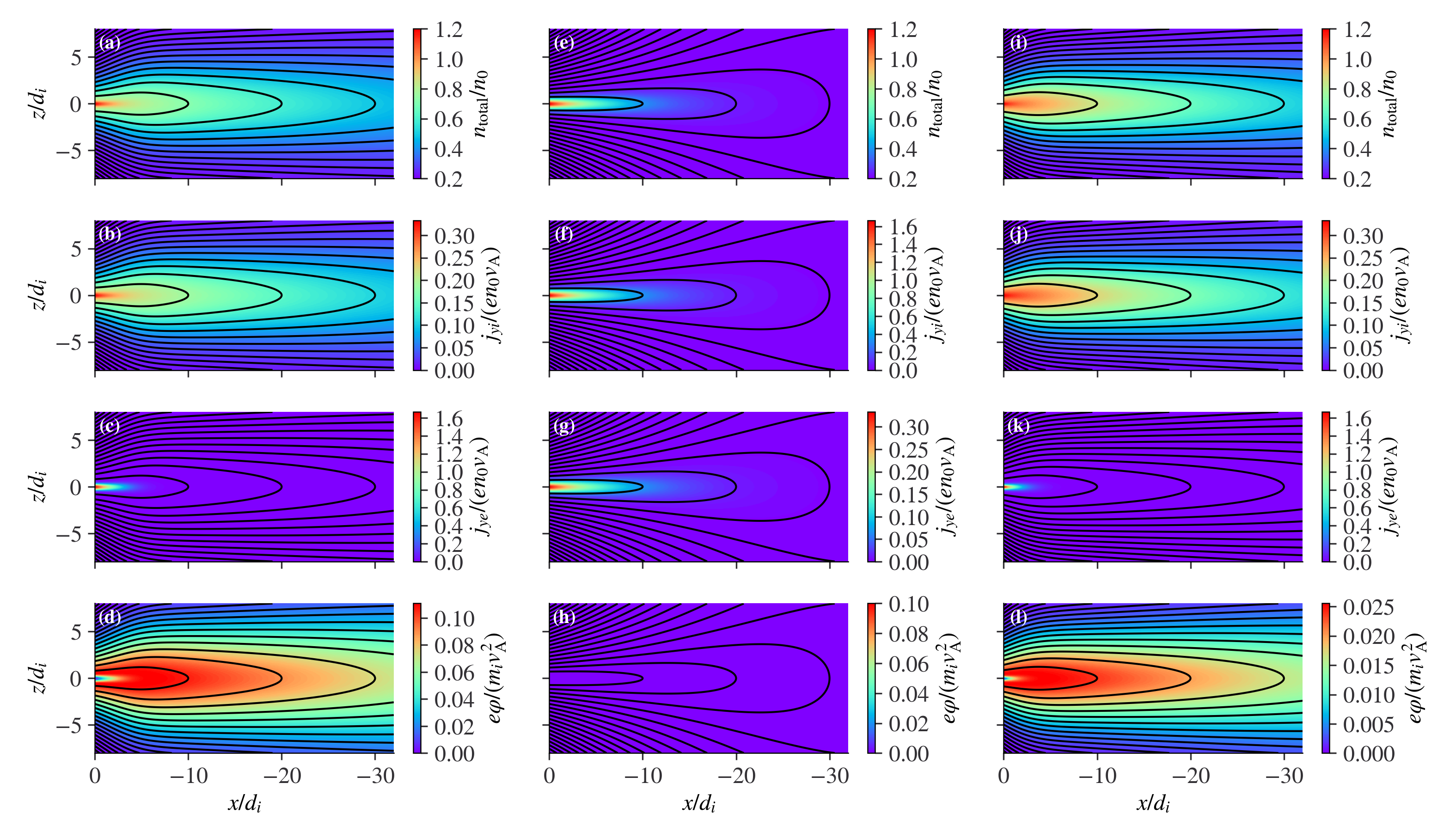}
    \caption{Configurations of polarized and nonpolarized Lembege-Pellat current sheets. The parameters used to set up the three current sheets are shown in Table \ref{tab:lembege-pellat-parameters}. (a), (b), (c), and (d) The polarized Lembege-Pellat current sheet with $T_{\alpha b} = T_{\alpha 0}$ ($\alpha = e, i$). (e), (f), (g), and (h) The nonpolarized Lembege-Pellat current sheet with $T_{\alpha b} = T_{\alpha 0}$ ($\alpha = e, i$). (i), (j), (k), and (l) The polarized Lembege-Pellat current sheet with $T_{\alpha b} = 0.2 T_{\alpha 0}$ ($\alpha = e, i$). The four rows from top to bottom display the total plasma density, the ion current density, the electron current density, and the electrostatic potential, respectively. The black lines in each panel stand for magnetic field lines (i.e., contours of $A_y$).}
    \label{fig:CS-configuration-comparison}
\end{figure*}

\begin{figure*}[tphb]
    \centering
    \includegraphics[width=\textwidth]{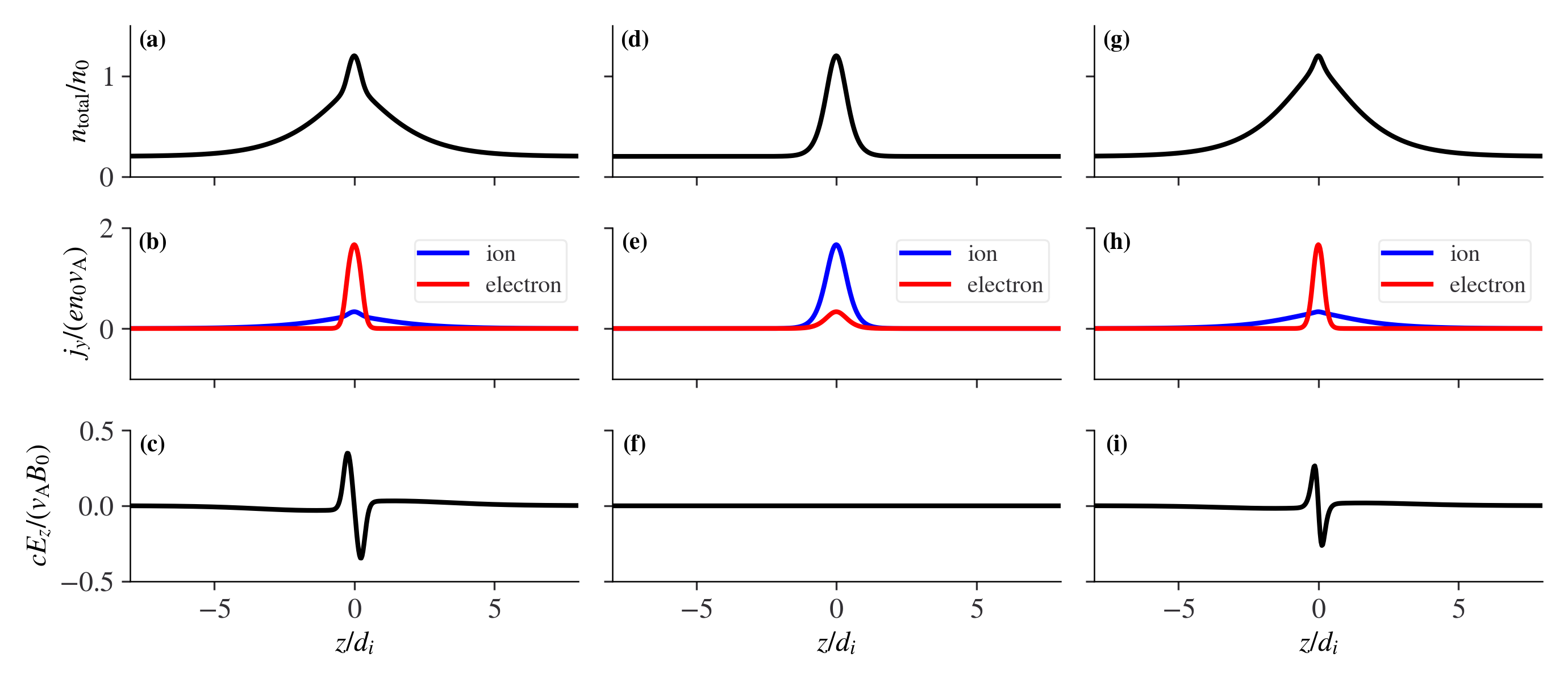}
    \caption{One-dimensional cuts of polarized and nonpolarized Lembege-Pellat current sheets from Figure \ref{fig:CS-configuration-comparison}. The three columns correspond to the three configurations in Figure \ref{fig:CS-configuration-comparison}, respectively. The three rows from top to bottom display the total plasma density, the ion and electron current densities, and the polarized electric field along the $z$-axis at $x = 0$, respectively.}
    \label{fig:CS-configuration-1d}
\end{figure*}

We study the reconnection properties of polarized current sheets and compare them with those of nonpolarized current sheets. Figure \ref{fig:CS-configuration-comparison} shows the initial conditions for three representative examples of current sheets with parameters from Table \ref{tab:lembege-pellat-parameters}. For easier comparison, the total current density peak is kept the same in the three examples. In the first example [Figures \ref{fig:CS-configuration-comparison}(a, b, c, d), Figures \ref{fig:CS-configuration-1d}(a, b, c)], we show a polarized current sheet with $- v_{De} T_{i0} / (v_{Di} T_{e0}) = 25$. The intense electron current sheet, which is embedded in a relatively weak ion current sheet, is representative of conditions in the near-Earth region (at around $12$-$15\, \mathrm{Re}$). In the second example [Figures \ref{fig:CS-configuration-comparison}(e, f, g, h), Figures \ref{fig:CS-configuration-1d}(d, e, f)], we show a nonpolarized current sheet with $- v_{De} T_{i0} / (v_{Di} T_{e0}) = 1$. The ion and electron current densities have the same spatial profiles, but a different multiplication factor (i.e., $\vert v_{Di} \vert > \vert v_{De} \vert$). The third current sheet example [Figures \ref{fig:CS-configuration-comparison}(i, j, k, l), Figures \ref{fig:CS-configuration-1d}(g, h, i)] is the same as the first one except that the ion and electron background temperatures are changed to $T_{\alpha b} = 0.2 T_{\alpha 0}$ ($\alpha = e, i$). Such a current sheet with decreased background temperature is intended to represent the thin current sheets observed in the mid-tail ($15$-$40\,\mathrm{Re}$ covered by Cluster and MMS orbits) \cite{Runov06,Artemyev17:jgr:THEMIS&Geotail,Lu19:jgr:cs}. Because of this modification relative to the first example, the profile of the ion current sheet in the third case is more extended farther along the current sheet.

Figures \ref{fig:CS-configuration-comparison} and \ref{fig:CS-configuration-1d} show that the current sheet polarization changes the plasma density profile. Note this effect occurs only if the current sheet is embedded into the background plasma sheet (as was noted before for 1D current sheets, see Ref.~\onlinecite{YL04}). Therefore, the plasma density redistribution (and corresponding reconfiguration of the current sheet due to change of the plasma pressure and magnetic field profiles) is the interplay of the electric field and background plasma: the dominance of electron currents ($|v_{De}|/T_{e}\gg v_{Di}/T_i$) requires a strong negative electrostatic potential $\varphi\sim v_{De}A_y/c$ (i.e., strong polarization electric field ${\bf E}=-\nabla \varphi$), and the background ion redistribution in this potential increases the plasma density around the equator, $n_i\sim \exp(-e\varphi/T_{ib})$. Note, the initial configuration of the polarized current sheet does not have field-aligned electric fields ($\varphi=\varphi(A_y)$ and $\nabla\varphi\cdot \left(\nabla\times{A_y{\bf e}_y}\right)=0$, see discussion in Refs.~\onlinecite{Birn04,Schindler12}), and thus the cross-field polarization field moves background ions without a field-aligned redistribution of electrons. This is drastically different the current sheet in the reconnection region, where field-aligned electric fields control the electron redistribution\cite{Egedal12,Egedal13}.

\begin{table*}[tphb]
	\centering
	\def\arraystretch{1}
	\begin{tabular}{c | c | c | c | c | c | c | c | c | c}
		\hline
		& $v_{Di}/v_\mathrm{A}$ & $v_{De}/v_\mathrm{A}$ & $T_{i0}/m_i v_\mathrm{A}^2$ & $T_{e0}/m_i v_\mathrm{A}^2$ & $T_{ib}/m_i v_\mathrm{A}^2$ & $T_{eb}/m_i v_\mathrm{A}^2$ & $n_b/n_0$ & $L_x/d_i$ & $L_z/d_i$ \\
		\hline
		Polarized & $1/3$ & $-5/3$ & $5/12$ & $1/12$ & $5/12$ & $1/12$ & $0.2$ & $32$ & $16$ \\
		Nonpolarized & $5/3$ & $-1/3$ & $5/12$ & $1/12$ & $5/12$ & $1/12$ & $0.2$ & $32$ & $16$ \\
		Polarized & $1/3$ & $-5/3$ & $5/12$ & $1/12$ & $1/12$ & $1/60$ & $0.2$ & $32$ & $16$
		\\
		\hline
	\end{tabular}
	\caption{\label{tab:lembege-pellat-parameters}Three sets of parameters for polarized and nonpolarized Lembege-Pellat current sheets. The velocities are normalized to the Alfv\'en velocity $v_\mathrm{A} = B_0 / \sqrt{4 \pi n_0 m_i}$, the temperatures are normalized to $m_i v_\mathrm{A}^2$, the densities are normalized to $n_0$, and the length is normalized to the ion inertial length $d_i$.}
\end{table*}

Initializing the current sheets as in Figure \ref{fig:CS-configuration-comparison}, we use a two-dimensional particle-in-cell (PIC) code \cite{pritchett2005newton,Lu18:pop} to study their evolution. Initialization obeying the density distributions as determined by the scalar and vector potentials in Equation \eqref{eq-quasi-neutrality-2}, which is non-trivial, is done using fast inverse transform sampling with function approximation by Chebyshev polynomials, as documented elsewhere \cite{an2022fast}. The relativistic equations of motion for ions and electrons are integrated in time using a leapfrog method. The electric and magnetic fields are advanced in time by integrating the time-dependent Maxwell's equations \cite{yee1966numerical} with $\mathbf{E}$ defined at integer time steps and $\mathbf{B}$ defined at half-integer time steps. The method uses a fully staggered grid (i.e., the Yee lattice) in which $\mathbf{E}$ and $\mathbf{J}$ are defined on the midpoints of the cell edges, whereas $\mathbf{B}$ is defined on the midpoints of the cell surfaces (e.g., see Figure 1 of Ref.~\onlinecite{wang19953d}). Charge conservation is guaranteed by adding a correction $\delta \mathbf{E}$ to the electric field \cite{langdon1976electromagnetic}, obtained by solving $\nabla \cdot \delta \mathbf{E} = 4 \pi \rho - \nabla \cdot \mathbf{E}$. The inclusion of $\delta \mathbf{E}$ has been proven to be important \cite{Pritchett96}; without it the results quickly become nonphysical.

Because the Lembege-Pellat current sheet \cite{LP82} with $B_z\ne 0$ is stable to spontaneous reconnection \cite{Pellat91}; so an external driver is needed to ensure instability.  At $z = \pm L_z / 2$, we apply a localized electric field of the form \cite{birn05,pritchett2005newton,Lu18:pop}
\begin{equation}\label{eq:driving-efield}
	E_{y, \mathrm{drive}} (x, t) = E_{y0} \cdot f(t) \cdot \sin^2\left(\frac{\pi x}{L_x}\right)
\end{equation}
to drive an inflow, where $f(t) = \tanh(\omega t) / \cosh^2(\omega t)$ describes the activation and deactivation of the applied electric field over a characteristic timescale $\omega^{-1} = 20\, \omega_{ci}^{-1}$ with $\omega_{ci} = B_0 e / m_i c$. New particles are injected into the system at $z = \pm L_z / 2$ at a rate determined by the background density and the $\mathbf{E} \times \mathbf{B}$ drift velocity. Particles crossing the $z$ boundaries are reflected back into the system.

At $x = -L_x$ and $x = 0$, particles are removed from the system at one cell inside the boundaries. New particles are continuously injected into the system with a vertical($z$) density profile identical to that of the initial current sheet. The velocities of these newly injected particles are distributed according to a one-sided Maxwellian \cite{aldrich1985particle}. Guard values of $B_z$ at the $x$ boundaries are used to advance $E_y$ in the Ampere's Law determined by a two-level time advancement method \cite{pritchett2005newton} in which the $B_z$ field is assumed to propagate at the speed of light. This method ensures that the magnetic flux can cross the $x$ boundaries.

The computational domain is $N_z \times N_x = 512 \times 1024$ in grid units with a grid scale $\delta = d_i / 32$. The Debye length is $\lambda_{De} = 0.46\, \delta$. The time step is $\Delta t = 0.001\, \omega_{ci}^{-1}$. The ion-to-electron mass ratio is $m_i / m_e = 100$. The ratio of the electron plasma frequency $\omega_{pe}$ to the electron cyclotron frequency $\omega_{ce}$ is $2$, which makes the normalized speed of light $c / v_\mathrm{A} = 20$ in our simulation. In the three examples, the density $n_0$ is represented by $1855$\,ppc (particles per cell), $3393$\,ppc, and $2827$\,ppc, respectively, which yields a total of $7.84 \times 10^8$ particles at the initial time for all. Our results are normalized to ion-based units: time to the inverse of the ion cyclotron frequency $\omega_{ci}$, length to the ion inertial length $d_i$, velocity to the Alfv\'en velocity $v_\mathrm{A}$, and energy to $m_i v_\mathrm{A}^2$.

\section{Simulation results\label{sec:simulation}}

\subsection{Reconnection onset}
\begin{figure*}[tphb]
    \centering
    \includegraphics[width=0.7\textwidth]{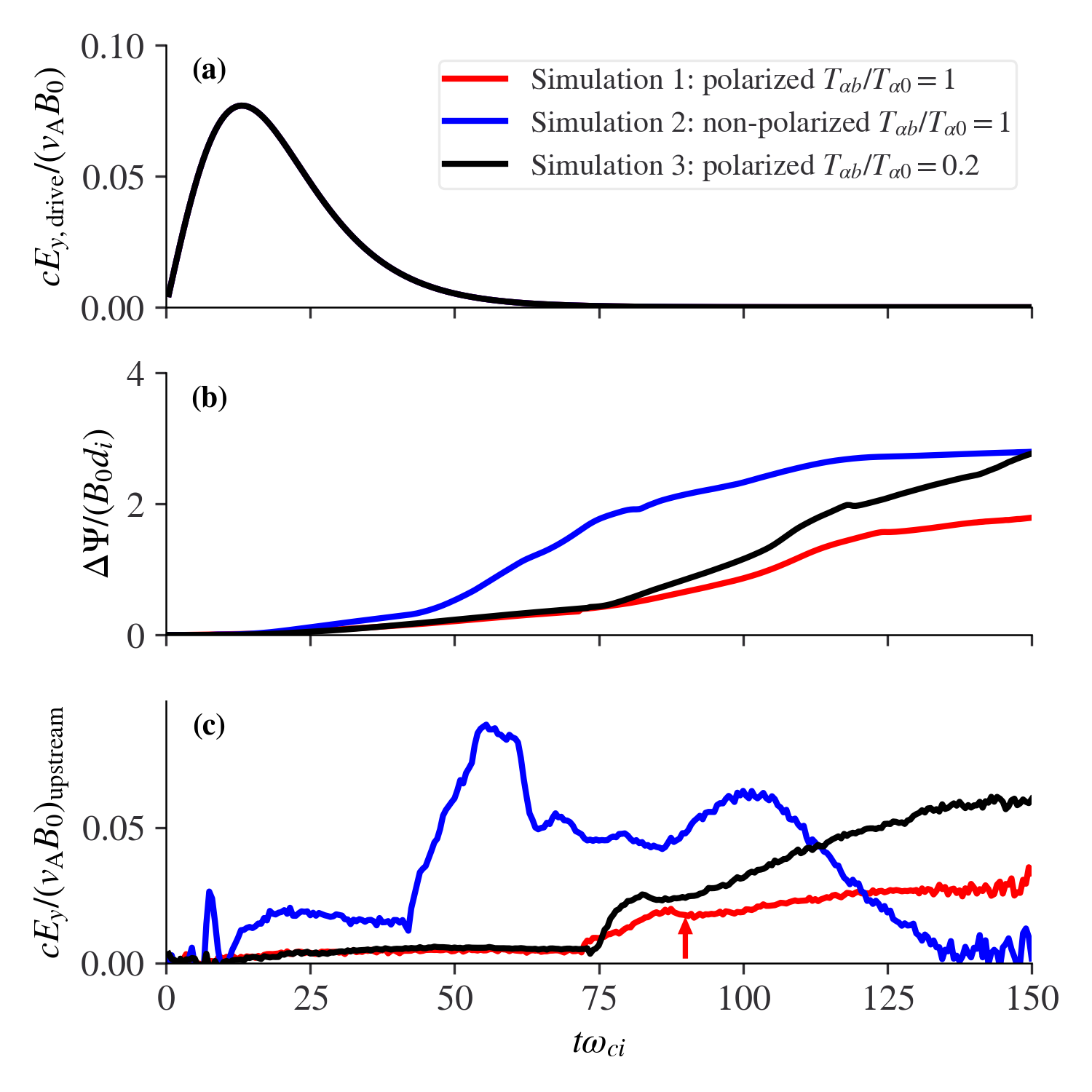}
    \caption{Evolution of magnetic reconnection for the three nominal current sheets. Here the three nominal current sheets refer to those in Figure \ref{fig:CS-configuration-comparison}. (a) The driving electric field at the $z$ boundaries for all the simulations. (b) The total reconnected magnetic flux. (c) The electric field at the reconnection site normalized to the instantaneous magnetic field magnitude and Alfv\'en speed in the upstream of reconnection region \cite{Karimabadi07}. This normalization ensures that any difference in reconnection electric field is caused by the properties of current sheet rather than the inflow conditions, and will be used throughout the paper. The electric field serves as a proxy for the reconnection rate $\mathrm{d} (\Delta\psi) / \mathrm{d} t$. Prior to the reconnection onset, the electric field is taken from the center of the computational domain. The arrow marks the time instant in Simulation $1$ at which we show the energy partition later in the Section \ref{subsec:partition}.}
    \label{fig:reconnection-rates}
\end{figure*}

Under the same driving electric field, the reconnected magnetic fluxes and reconnection electric fields in the three nominal current sheets are shown in Figure \ref{fig:reconnection-rates}. We refer to the simulations corresponding to the three examples of current sheets in Figure \ref{fig:CS-configuration-comparison} as Simulations $1$, $2$ and $3$. In each simulation, the reconnection onset is preceded by a driven phase, during which magnetic flux in the simulation volume increases because of the input flux from the driver electric field at the top and bottom boundaries.

Both the total reconnected magnetic flux and the maximum of the reconnection rate are lower in the polarized current sheet in Simulation $1$ than in the nonpolarized current sheet in Simulation $2$. This is likely because in the current sheet embedded into the background plasma the polarization changes plasma density profile, which results in thicker profiles of plasma pressure and magnetic field in the intermediate $x$-range (see Figure \ref{fig:CS-configuration-comparison}). Thus, the cold plasma density redistribution by polarization electric field changes the current sheet configuration that controls the reconnection rate \cite{Liu17:reconnection_rate}. In addition, the reconnection onset in the polarized current sheet (Simulation $1$) occurs later than that in the nonpolarized current sheet (Simulation $2$). These results are consistent with the findings of a previous study \cite{Lu20:pop} that compared reconnection in polarized and nonpolarized Harris current sheets in 1D.

\begin{figure*}[tphb]
    \centering
    \includegraphics[width=\textwidth]{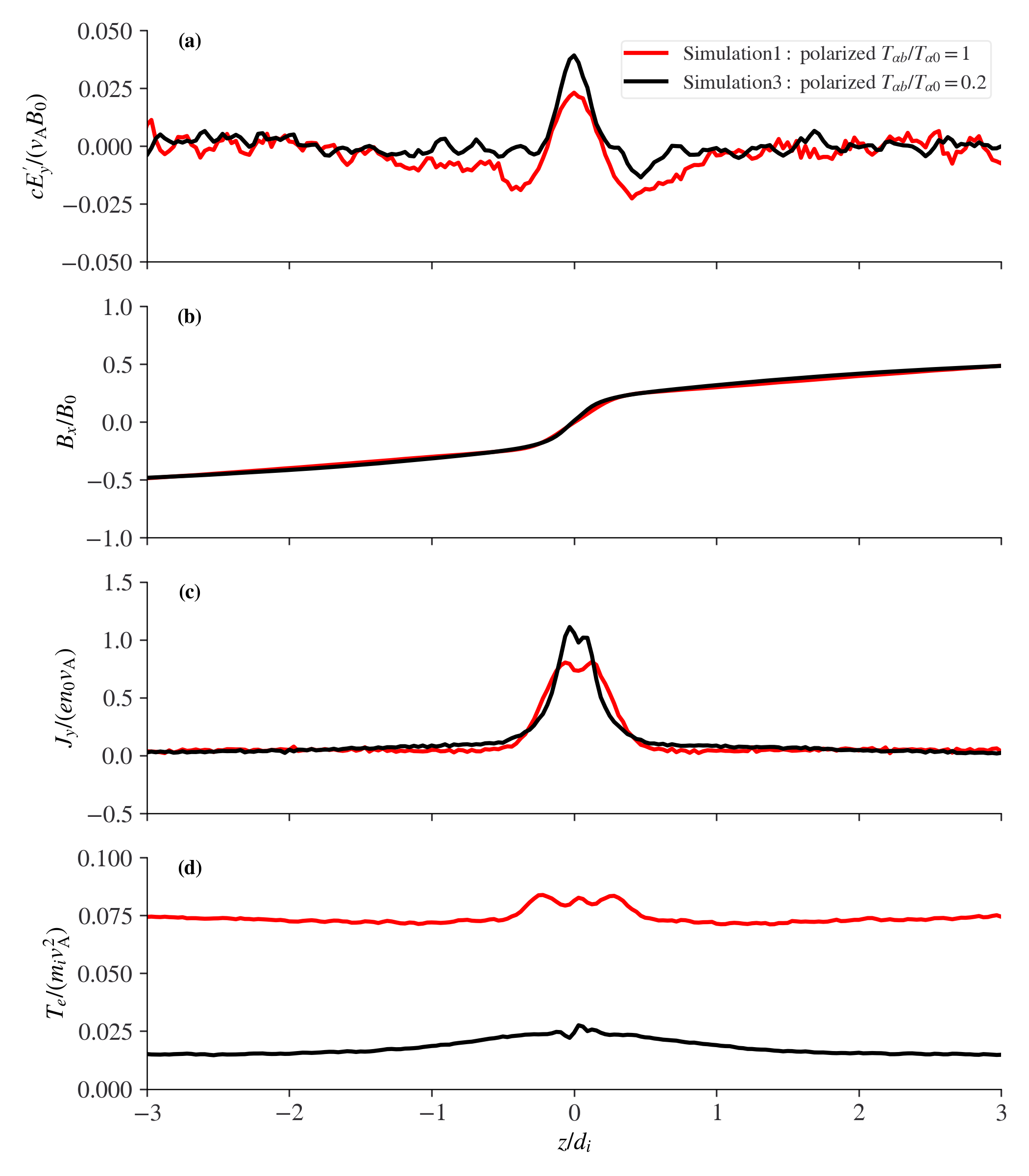}
    \caption{Profiles of non-ideal electric field $E_y^\prime$ (a), \blue{magnetic field $B_x$ (b)}, current density $J_y$ (c), and electron temperature $T_e$ (d) at the reconnection site along the $z$-axis. The results for Simulations $1$ and $3$ are both taken at $t = 120\ \omega_{ci}^{-1}$ and are shown in red and black lines, respectively. The non-ideal electric field is calculated as $E_y^\prime = E_y + (\mathbf{v}_e \times \mathbf{B})_y / c$, where $\mathbf{v}_e$ is the electron flow velocity.}
    \label{fig:eyfield-zscale}
\end{figure*}

\blue{Decreasing the background temperature from that in Simulation $1$ ($T_{\alpha b} = T_{\alpha 0}$) to that in Simulation $3$ ($T_{\alpha b} = 0.2 T_{\alpha 0}$), results in an increase in the reconnection rate and the total reconnected magnetic flux. 
To check possible mechanisms responsible for the reconnection rate variation with the backgound temperature, we plot main current sheet characteristics for Simulations $1$ and $3$ in Figure \ref{fig:eyfield-zscale}. The scale length of the current sheet $L_z = \vert\mathrm{d} \ln B_x / \mathrm{d} z\vert^{-1}\approx c B_x/ (4\pi J_y)$ is slightly reduced in Simulation $3$: Figure \ref{fig:eyfield-zscale} shows slightly thinner profile of $J_y$ and higher $J_y$ magnitude for the current sheet with lower background electron temperature. It is not clear if such difference in the current sheet configuration may result in a significant change of charged electron dynamics leading to the change of the reconnection rate. Thus, further studies are needed to understand how the electron kinetics in the diffusion region \cite{Hesse99} respond to the upstream temperature.}

\begin{figure*}[tphb]
    \centering
    \includegraphics[width=0.8\textwidth]{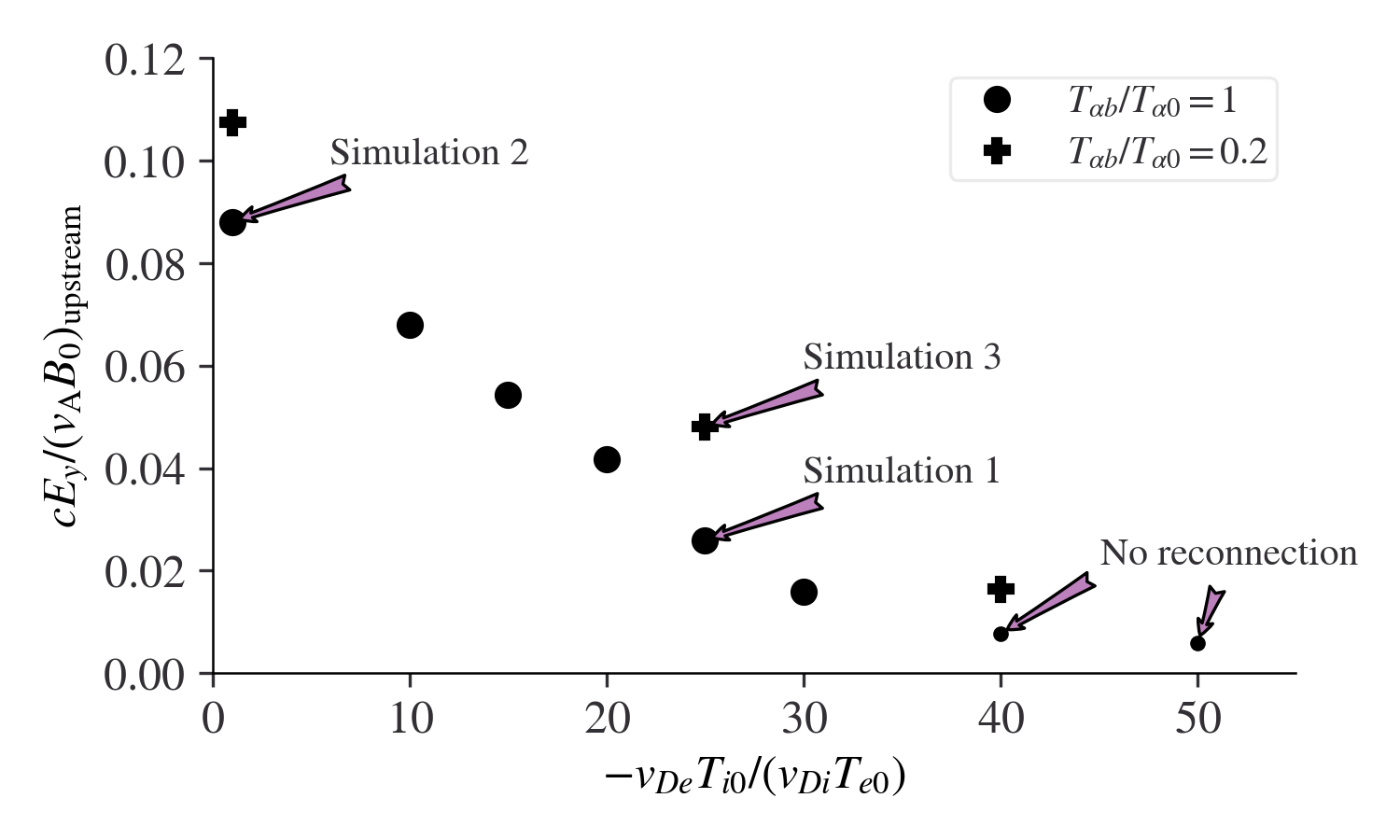}
    \caption{Maximum reconnection rate as a function of the polarization parameter $\lambda = - v_{De} T_{i0} / (v_{Di} T_{e0})$. The electric field normalization is the same as Figure \ref{fig:reconnection-rates}(c). The runs with $T_{\alpha b} = T_{\alpha 0}$ ($\alpha = e, i$) are represented by dots. The runs with decreased background temperatures $T_{\alpha b} = 0.2 T_{\alpha 0}$ ($\alpha = e, i$) are represented by the plus signs.}
    \label{fig:parameter_scan}
\end{figure*}

To reveal the effect of current sheet polarization on reconnection rate, we perform a series of runs by varying the polarization parameter $\lambda = - v_{De} T_{i0} / (v_{Di} T_{e0})$ as shown in Figure \ref{fig:parameter_scan}, where $\lambda = 1$ represents nonpolarized current sheets. As $\lambda$ increases, current sheets become more strongly polarized and hence more electron-dominant. In this parametric study, the total current density peak is kept constant (i.e., $(v_{Di} - v_{De}) / v_\mathrm{A} = 2$), and so is the electric field driver at the $z$ boundaries [see Figure \ref{fig:reconnection-rates}(a)]. Because of the progressively thicker plasma density (and thus, plasma pressure) profiles in increasingly more electron-dominant current sheets, current sheet configuration corresponds to less spatially localized current density (see Figure \ref{fig:CS-configuration-comparison}) and to smaller reconnection rate, which decreases linearly with $\lambda$, until no reconnection occurs at large enough $\lambda$ ($\lambda = 40$ and $50$ in Figure \ref{fig:parameter_scan}). \blue{Moreover, Figure \ref{fig:parameter_scan} confirms that the reconnection rate increases as the background temperature is decreased, although the precise mechanism for such increase is to be investigated in the next study.}

These simulation results indicate that the polarization may make thin current sheets more stable to magnetic reconnection, i.e., a stronger external driver is needed to trigger the reconnection in an increasingly polarized current sheet.
The intensity of the current sheet polarization varies along the magnetotail (with the radial distance) \cite{Artemyev16:jgr:ex}, and less polarized (containing significant ion currents) current sheets are predominantly observed at larger radial distances \cite{Vasko15:jgr:cs,Xu18:artemis_cs}, whereas current sheet polarization is stronger closer to the Earth \cite{Runov06,Artemyev09:angeo}.
Therefore, addressing one of the most important questions in magnetotail dynamics, what controls the reconnection onset location \cite{Baker96,Baumjohann00,Genestreti14}, requires proper estimation of the current sheet polarization and the background plasma characteristics at different radial distances from the Earth.

\subsection{Energy partition\label{subsec:partition}}
We now turn our attention to the partition of the initial magnetic energy of the reconnecting plasma into the final, outflow energy (e.g., bulk flows, heating). We examine the outflow energy transport in polarized current sheets, and compare its components with those for nonpolarized current sheets. 

The total energy density has three main components: the electric and magnetic energy densities $(E^2 + B^2) / (8 \pi)$ (note $E^2 \ll B^2$ for magnetospheric conditions), the sum of the kinetic energy density of bulk flows of all species $K_\alpha = \frac{1}{2}n_\alpha m_\alpha v_\alpha^2$, and the sum of the thermal energy density of all species $u_\alpha = \frac{1}{2} \mathrm{Tr}(\mathbf{P}_\alpha)$. Here $\mathbf{P}_\alpha = m_s \int (\mathbf{w} - \mathbf{v}_\alpha) (\mathbf{w} - \mathbf{v}_\alpha) f_\alpha \mathrm{d}^3\mathbf{w}$ is the pressure tensor of species $\alpha$; and $\mathrm{Tr}(\cdot)$ is the trace operator. In the Eulerian formalism, energy conservation reads \cite{Birn&Hesse05,aunai2011energy,Lu18:pop}
\begin{eqnarray}
\frac{\partial}{\partial t}\left(\frac{B^2 + E^2}{8 \pi}\right) + \nabla \cdot \mathbf{\Pi} = -\mathbf{J} \cdot \mathbf{E}, \label{eq:magnetic-energy}\\
\frac{\partial}{\partial t} K_\alpha + \nabla \cdot (K_\alpha \mathbf{v}_\alpha) = S_{k\alpha}, \label{eq:flow-energy}\\
\frac{\partial}{\partial t} u_\alpha + \nabla \cdot (\mathbf{H}_\alpha + \mathbf{Q}_\alpha) = S_{u\alpha}. \label{eq:thermal-energy}
\end{eqnarray}
In each equation, the change in local energy density per unit time and the divergence of the local energy flux are balanced by a source term. The energy fluxes describe the energy transport while the source terms describe the conversion between different forms of energy. Specifically, magnetic energy is transported by the Poynting flux $\mathbf{\Pi} = \frac{c}{4 \pi} \mathbf{E} \times \mathbf{B}$; kinetic energy is transported by its flux term $K_\alpha \mathbf{v}_\alpha$; thermal energy is transported by the sum of the enthalpy flux $\mathbf{H}_\alpha = u_\alpha \mathbf{v}_\alpha + \mathbf{P}_\alpha \cdot \mathbf{v}_\alpha$ and the heat flux $\mathbf{Q}_\alpha = \frac{1}{2} m_\alpha \int (\mathbf{w} - \mathbf{v}_\alpha)^2 (\mathbf{w} - \mathbf{v}_\alpha) f_\alpha \mathrm{d}^3\mathbf{w}$. The source terms for the kinetic and thermal energies are $S_{k\alpha} = \mathbf{J}_\alpha \cdot \mathbf{E} - (\nabla \cdot \mathbf{P}_\alpha) \cdot \mathbf{v}_\alpha$ and $S_{u\alpha} = (\nabla \cdot \mathbf{P}_\alpha) \cdot \mathbf{v}_\alpha$, respectively. The sum of all the source terms is zero, i.e. $-\mathbf{J} \cdot \mathbf{E} + \sum_\alpha (S_{k\alpha} + S_{u\alpha}) = 0$, because of the conservation of total energy.

\begin{figure*}[tphb]
    \centering
    \includegraphics[width=\textwidth]{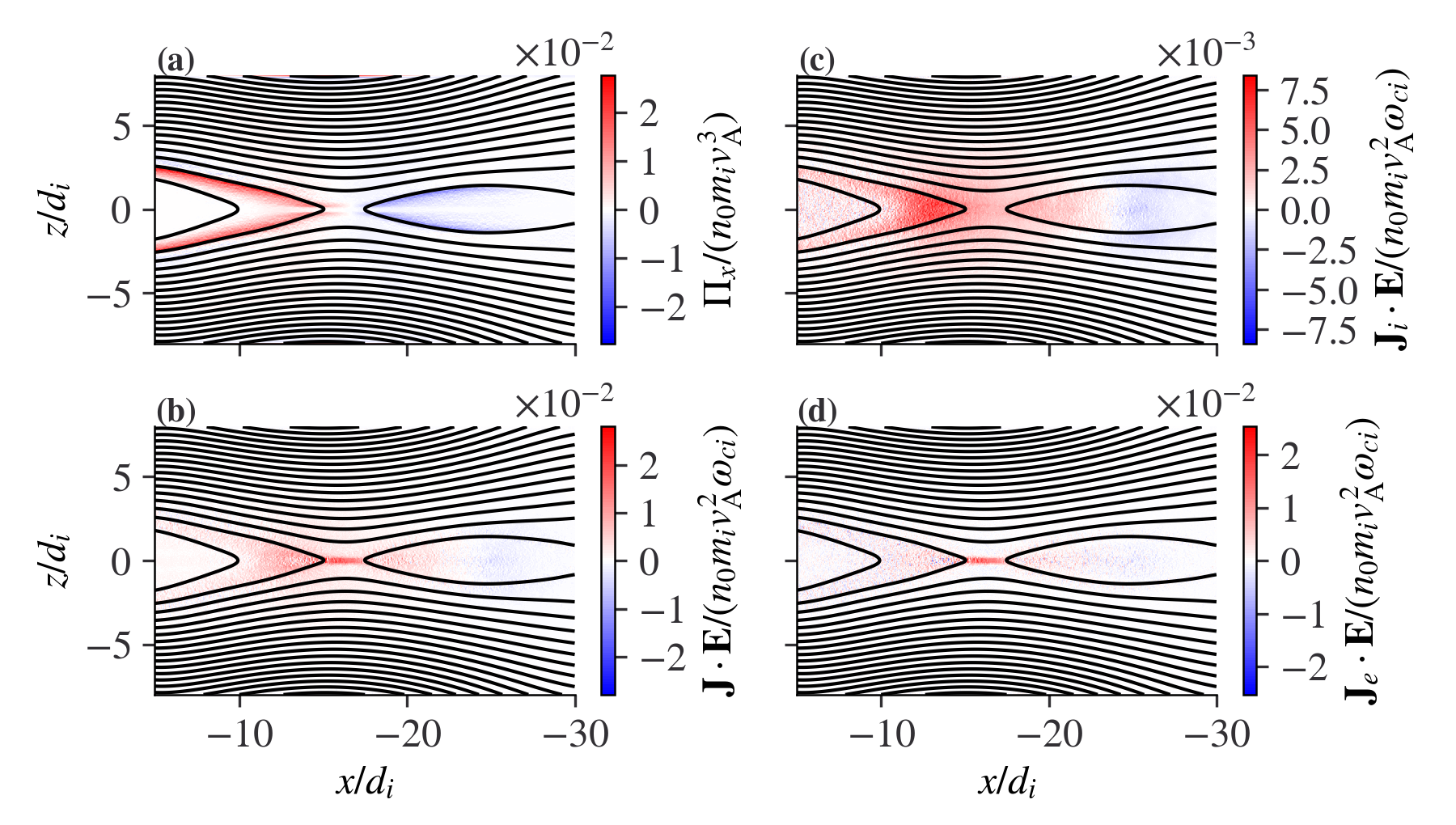}
    \caption{Conversion of magnetic energy in the polarized current sheet at $t = 90\,\omega_{ci}^{-1}$ in Simulation $1$. (a) Poynting flux in the $x$ direction. (b), (c), (d) The source terms $\mathbf{J} \cdot \mathbf{E}$, $\mathbf{J}_i \cdot \mathbf{E}$ and $\mathbf{J}_e \cdot \mathbf{E}$.}
    \label{fig:poynting}
\end{figure*}

In Figure \ref{fig:poynting} we show an example, the energy conversion near the maximum reconnection rate (at $t = 90\,\omega_{ci}^{-1}$) in the polarized current sheet of Simulation $1$, and then compare the energy transport for polarized and nonpolarized current sheets in the three representative simulations.

The overall structure of the dissipation region is shown in Figure \ref{fig:poynting}(b). Near the X-line ($-17.5 < x / d_i < -15$ and $z \sim 0$), unmagnetized electrons are accelerated by $E_y$ and form a thin current layer (with thickness close to the local electron inertial length \cite{Hesse99}), so the released magnetic energy is converted to electron energy [Figure \ref{fig:poynting}(d)]. In a broader region around the X-line ($-22 < x / d_i < -10$ and $-2 < z/d_i < 2$), where ions are still unmagnetized, ion acceleration by $E_y$ provides another means of magnetic energy conversion and dissipation, evidenced by $\mathbf{J}_i \cdot \mathbf{E}$ [Figure \ref{fig:poynting}(c)]. Part of the magnetic energy released at the reconnection site is transported via Poynting flux to the downstream region [Figure \ref{fig:poynting}(a)]. This Poynting flux provides a free energy source for particle energization and $B_z$ enhancement far away from the X-line in dipolarization fronts, supported by spacecraft observations \cite{Zhou12:acceleration,Angelopoulos13, Runov15} and numerical simulations \cite{Sitnov09,Birn14,Birn15,Lu16:acceleration,Pritchett&Runov17}. In addition, as previously noted in nonpolarized current sheets, there is an earthward-tailward asymmetry of $\Pi_x$ and $\mathbf{J} \cdot \mathbf{E}$, with more magnetic energy released toward the Earth \cite{Lu18:pop}. This asymmetry is mainly caused by the ion pressure $x$-gradient balancing the magnetic curvature force in the initial equilibrium, both arising from the finite $B_z$ (see discussions in Refs.~\onlinecite{Pritchett05:driven,Birn&Hesse14,Lu18:pop,Sitnov21:grl}); $\mathbf{J}_e \cdot \mathbf{E}$, however, is roughly symmetric with the reconnection site.
 
\begin{figure*}[tphb]
    \centering
    \includegraphics[width=\textwidth]{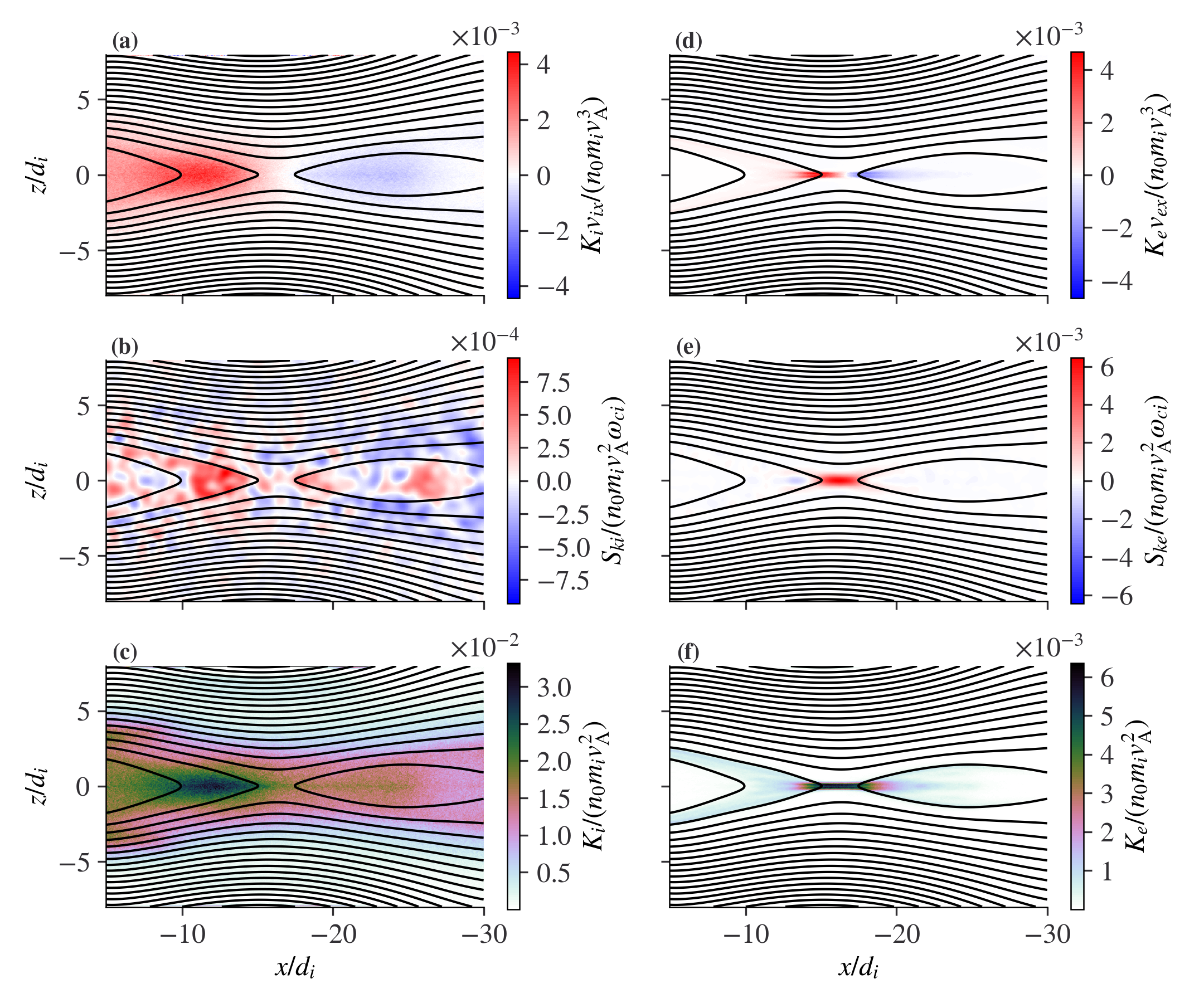}
    \caption{Physical quantities related to Equation \eqref{eq:flow-energy} for the conservation of bulk kinetic energy in Simulation $1$. The snapshots are taken at $t = 90\,\omega_{ci}^{-1}$. (a), (b), and (c) are the ion kinetic energy flux, the source term $S_{ki}$ and the ion kinetic energy density, respectively. (d), (e), and (f) are the electron kinetic energy flux, the source term $S_{ke}$ and the electron kinetic energy density, respectively. In calculating $S_{ki}$ and $S_{ke}$, we applied a Gaussian filter with a standard deviation $\sigma = 0.25\, d_i$ to the pressure tensors $\mathbf{P}_i$ and $\mathbf{P}_e$ to reduce noise in the gradient terms $\nabla \cdot \mathbf{P}_i$ and $\nabla \cdot \mathbf{P}_e$.}
    \label{fig:bulkflow}
\end{figure*}

\begin{figure*}[tphb]
    \centering
    \includegraphics[width=\textwidth]{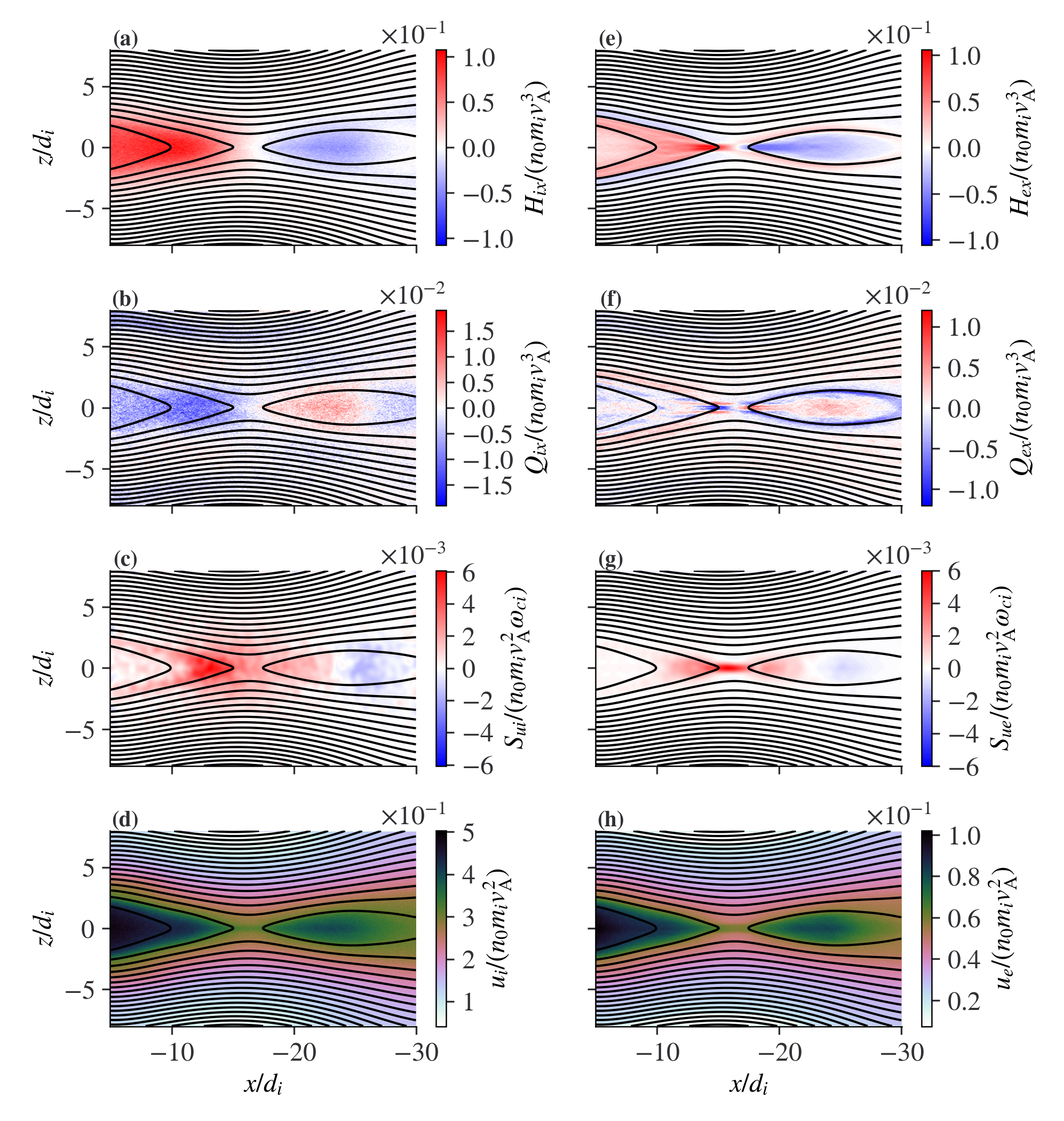}
    \caption{Physical quantities related to Equation \eqref{eq:thermal-energy} for the conservation of thermal energy in Simulation $1$. The snapshots are taken at $t = 90\,\omega_{ci}^{-1}$. (a), (b), (c), (d) are the ion enthalpy flux, the ion heat flux, the source term $S_{ui}$ and the ion thermal energy density, respectively. (e), (f), (g), (h) are the electron enthalpy flux, the electron heat flux, the source term $S_{ue}$ and the electron thermal energy density, respectively. The same Gaussian filtering technique as that in Figures \ref{fig:bulkflow}(b) and \ref{fig:bulkflow}(e) was used to reduce noise in $S_{ui}$ and $S_{ue}$.}
    \label{fig:thermal}
\end{figure*}

We next investigate the detailed process of particle acceleration and heating by analyzing the relevant quantities from Equations \eqref{eq:flow-energy} and \eqref{eq:thermal-energy}, as shown in Figures \ref{fig:bulkflow} and \ref{fig:thermal}. The total particle energization rate $\mathbf{J}_\alpha \cdot \mathbf{E}$ is split into the bulk acceleration rate $S_{k \alpha} = \mathbf{J}_\alpha \cdot \mathbf{E} - (\nabla \cdot \mathbf{P}_\alpha) \cdot \mathbf{v}_\alpha$ and the heating rate $S_{u \alpha} = (\nabla \cdot \mathbf{P}_\alpha) \cdot \mathbf{v}_\alpha$. The ion bulk acceleration rate [Figure \ref{fig:bulkflow}(b)] in the ion diffusion region is shown to be much smaller than the ion heating rate [Figure \ref{fig:thermal}(c)], and therefore the latter dominates the energy conversion process. The electron bulk acceleration rate is, on the other hand, comparable to the electron heating rate in the electron diffusion region [Figures \ref{fig:bulkflow}(e) and \ref{fig:thermal}(g), respectively]. As both species flow out of their respective diffusion regions and become magnetized, the bulk acceleration rates [Figures \ref{fig:bulkflow}(b) and \ref{fig:bulkflow}(e)] transition from being away from the reconnection site to being towards it. Some of the bulk kinetic energy from this deceleration is converted to thermal energy, as evidenced by the enhanced heating rates in the respective deceleration regions of ions and electrons [Figures \ref{fig:bulkflow}(c), \ref{fig:bulkflow}(f), \ref{fig:thermal}(c) and \ref{fig:thermal}(g)]. Again, an earthward-tailward asymmetry arises in the bulk acceleration and heating rates of ions, which is ultimately attributed to the asymmetry in $\mathbf{J}_i \cdot \mathbf{E}$ [Figure \ref{fig:poynting}(c)]. In contrast, the bulk acceleration and heating rates of electrons are roughly symmetric with respect to the reconnection site.

Kinetic and thermal energies are transported earthward and tailward from the reconnection site through various energy fluxes in the $x$ direction, including the kinetic energy flux $K_{\alpha} v_{\alpha x}$, the enthalpy flux $H_{\alpha x}$ and the heat flux $Q_{\alpha x}$. The sign of $K_{\alpha} v_{\alpha x}$ [Figures \ref{fig:bulkflow}(a) and \ref{fig:bulkflow}(d)] must be the same as that of the flow velocity $v_{\alpha x}$. However, the sign of $H_{\alpha x} = u_\alpha v_{\alpha x} + (\mathbf{P}_\alpha \cdot \mathbf{v})_x$ [Figures \ref{fig:thermal}(a) and \ref{fig:thermal}(e)], although not required to be, is also the same as that of $v_{\alpha x}$, implying that the contribution from the diagonal part of the pressure tensor $u_\alpha v_{\alpha x} + P_{\alpha, xx} v_{\alpha x}$ dominates over that from the off-diagonal part $P_{\alpha, xy} v_{\alpha y} + P_{\alpha, xz} v_{\alpha z}$. The heat flux $Q_{\alpha x}$, being a third-order moment of the distribution function, has more complex spatial structures [Figures \ref{fig:thermal}(b) and \ref{fig:thermal}(f)] and is not well organized by the flow pattern of $v_{\alpha x}$. Because of the very different widths between the ion and electron outflows (i.e., $v_{ix}$ and $v_{ex}$) as well as the complication by the electron beams along the separatrix, the regions with appreciable electron energy fluxes are more concentrated and show finer structures than the spatial distributions of ion energy fluxes.

\begin{figure*}[tphb]
    \centering
    \includegraphics[width=\textwidth]{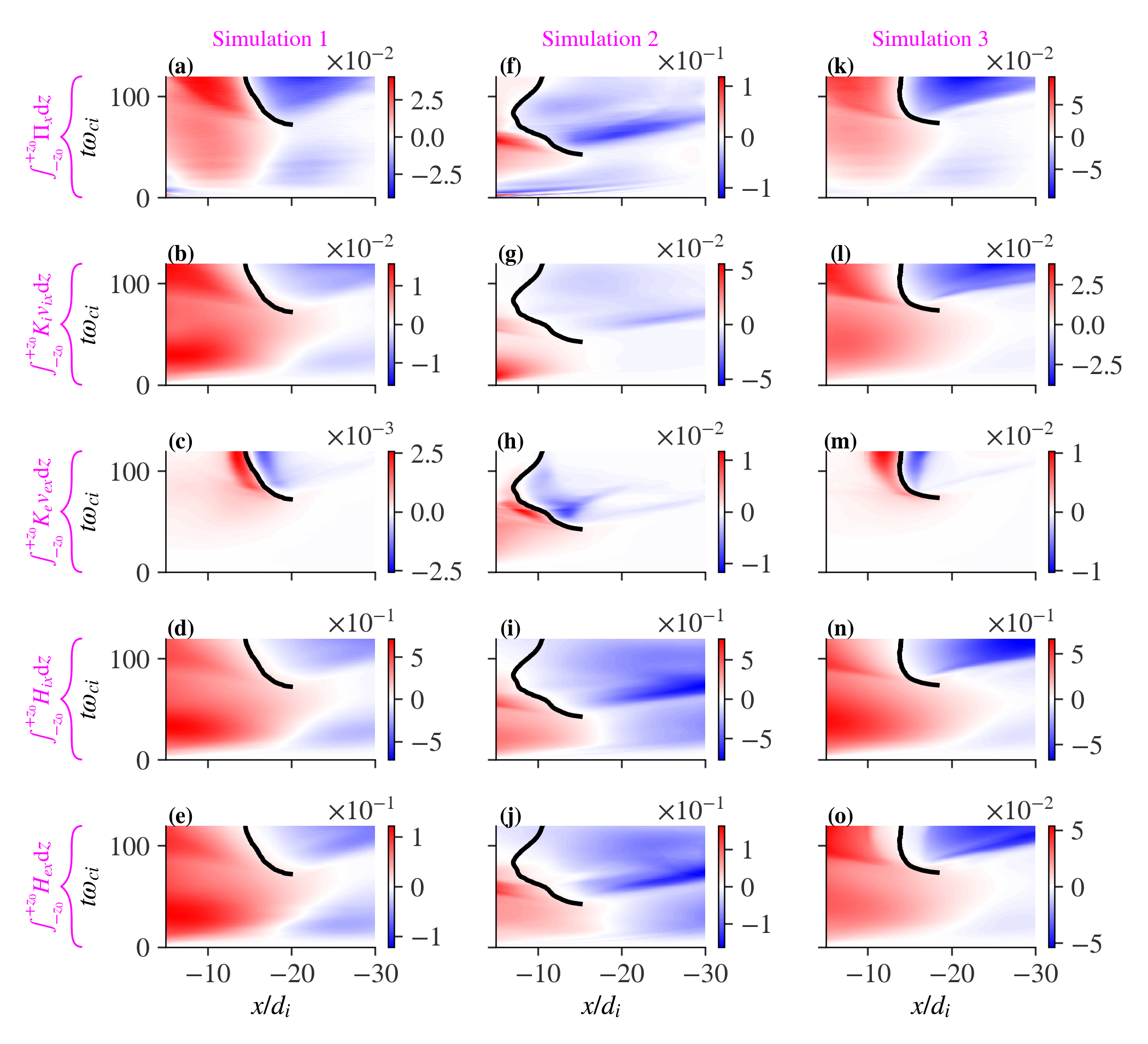}
    \caption{Spatiotemporal evolution of the line-integrated energy flux. The three columns from left to right are for Simulations $1$, $2$ and $3$, respectively. The five rows from top to bottom show $\int_{-z_0}^{+z_0} \Pi_x\, \mathrm{d}z$, $\int_{-z_0}^{+z_0} K_i v_{ix}\, \mathrm{d}z$, $\int_{-z_0}^{+z_0} K_e v_{ex}\, \mathrm{d}z$, $\int_{-z_0}^{+z_0} H_{ix}\, \mathrm{d}z$ and $\int_{-z_0}^{+z_0} H_{ex}\, \mathrm{d}z$, respectively. Here the integration limit is $z_0 = \frac{L_z}{2} - \Delta z$ with $\Delta z = 2 d_i$, where $\Delta z$ is introduced to avoid boundary effects. The black solid line in each panel represents the $x$ position of the reconnection site (i.e., $B_z = 0$) as a function of time. The starting point of this line indicates the reconnection onset.}
    \label{fig:enefluxx}
\end{figure*}

\begin{figure*}[tphb]
    \centering
    \includegraphics[width=\textwidth]{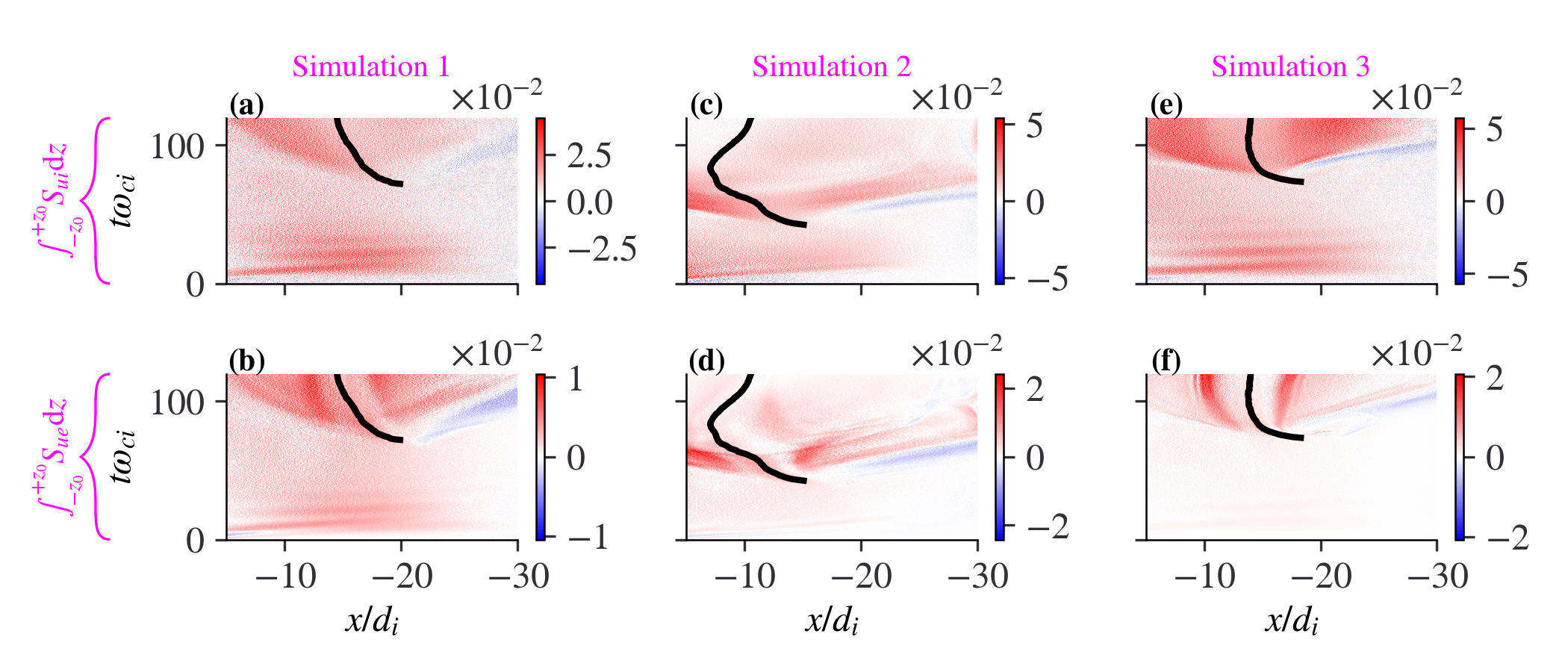}
    \caption{Spatiotemporal evolution of the line-integrated plasma heating rate. The format is the same as that of Figure \ref{fig:enefluxx}. The two rows from top to bottom show the integrated ion and electron heating rates, i.e., $\int_{-z_0}^{+z_0} S_{ui}\, \mathrm{d}z$ and $\int_{-z_0}^{+z_0} S_{ue}\, \mathrm{d}z$, respectively.}
    \label{fig:local-heating-rate}
\end{figure*}

Lastly, we compare the relative ordering of different forms of energy flux in the three representative simulations. To achieve this, we integrate each energy flux term along the $z$ direction at any given $x$ and compare the line-integrated energy flux as function of position $x$ and time $t$. The results are shown in Figure \ref{fig:enefluxx}. In all representative simulations, the energy flux is mostly in the form of ion enthalpy flux, followed by electron enthalpy flux, Poynting flux, ion kinetic energy flux and electron kinetic energy flux, i.e., $H_{ix} > H_{ex} > \Pi_x > K_{i} v_{ix} > K_{e} v_{ex}$. The heat flux is negligible relative to the enthalpy flux. This ordering is consistent with previous spacecraft observations \cite{Eastwood13} and numerical simulations for nonpolarized current sheets \cite{Birn&Hesse14}. The directions of energy flux terms are well organized by the location of the X-line and except for the heat fluxes (not shown) are directed away from the X-line. Before reconnection onset, externally driven inflows result in outflows because of the mass continuity, which carries significant amounts of $H_{ix}$, $H_{ex}$, $\Pi_x$ and $K_i v_{ix}$ even in comparison with those after the reconnection onset. Comparing Simulations $1$ and $2$, it is seen that that the Poynting and kinetic energy fluxes in the polarized current sheet are smaller than those in the nonpolarized current sheet, whereas the ion and electron enthalpy fluxes are comparable in the two types of current sheets. This indicates that the bulk acceleration efficiency is significantly decreased by the current sheet polarization, yet the plasma heating efficiency might be irrelevant to the current sheet polarization. With the lower background temperatures in Simulation $3$ than in Simulation $1$, we observe increases in the Poynting and kinetic energy fluxes, comparable ion enthalpy flux, and a decrease in the electron enthalpy flux. These indicates that although the bulk acceleration efficiency is enhanced by the lower background temperature, the plasma heating efficiency (predominantly the ion heating efficiency) is relatively unchanged. Since appreciable enthalpy fluxes are caused by plasma convection from the inflow region to the outflow region even before the reconnection onset [Figures \ref{fig:enefluxx}(d-e), \ref{fig:enefluxx}(i-j), \ref{fig:enefluxx}(n-o)], it is necessary to compare the integrated local heating rate $S_{u \alpha}$ to assess the impact of polarization on plasma heating by reconnection itself. The results are shown in Figure \ref{fig:local-heating-rate}. Significant plasma heating occurs around the reconnection site after the reconnection onset in the three simulations. We confirm that the plasma heating rate (predominantly ion heating rate) is relatively unchanged from polarized to nonpolarized current sheets. We also notice the recent MMS results \cite{eastwood2020energy} on the importance of electron enthalpy flux in the out-of-plane direction in the energy transport near electron diffusion region in the magnetopause reconnection, which is out of the scope of our study.

\section{Discussion and Conclusions\label{sec:discussion}}
Our simulations of magnetic reconnection in 2D polarized current sheet confirm results previously obtained for the 1D Harris current sheet model \cite{Lu20:pop}: such current sheets characterized by weak initial ion current and strong electron current attain smaller reconnection rates than nonpolarized, ion-dominant current sheets. Because 2D current sheets with $B_z\ne 0$ are stable to the tearing mode \cite{Pellat91,Quest96,Sitnov02}, we find that a stronger external driver is needed to drive the thinning and reconnection of polarized current sheets. In the Earth's magnetotail, such polarization is associated with current sheet thinning during the substorm growth phase \cite{Pritchett&Coroniti95,Hesse98}, and such polarized current sheet with strong electron currents are observed before reconnection onset \cite{Wang18:mms, Lu19:jgr:cs,Richard21}. But the current sheet thinning is a spatially localized process: ion-dominant thin current sheets may exist at small radial distances \cite{Artemyev16:jgr:ex} and very large (lunar orbit) radial distances \cite{Xu18:artemis_cs}, whereas mid-tail current sheets are characterized by strong electron currents \cite{Runov06,Artemyev09:angeo,Vasko15:jgr:cs, Hubbert21}. These near-Earth thin current sheets have larger current density and more free (upstream) magnetic field energy than the mid-tail current sheets \cite{Yushkov21}, but are very stable. Our study shows that stabilization of current sheet by stronger polarization electric fields may explain why near-Earth ion-scale thin current sheets \cite{Artemyev16:jgr:thinning} are stable to magnetic reconnection, which is consistent with observations that reconnection almost never occurs there. Therefore, the location of the reconnection onset may be determined not only by the strongest current density, but also by the interplay between the current density increase and current sheet polarization.


To conclude, we consider the effects of the current sheet polarization on magnetic reconnection in a current sheet with $B_z \neq 0$. This configuration is not only typical of the Earth's magnetotail, but can also be found in other planetary magnetotails \cite{Jackman14}. A current sheet with a finite $B_z$ (either with or without polarization) is stable to spontaneous reconnection (except for spatially localized $B_z$ peaks, see Refs.~\onlinecite{Sitnov14, Bessho&Bhattacharjee14, Pritchett15:pop, Merkin15,Birn18}) and thus requires external driving to trigger reconnection. We have found that the required external driving is stronger in the polarized current sheet. Furthermore, polarization makes the plasma density (and plasma pressure) profiles thicker in the equilibrium current sheet, which eventually leads to lower reconnection rate and weaker post-reconnection energy flows in the polarized current sheet. Our results demonstrate that a subtle kinetic effect, current sheet polarization, may significantly alter the properties of reconnection with global consequences for magnetospheric energy conversion and transport.

\begin{acknowledgements}
The work was supported by NASA awards 80NSSC18K1122, 80NSSC20K1788, and NAS5-02099. We would like to acknowledge high-performance computing support from Cheyenne (\url{doi:10.5065/D6RX99HX}) provided by NCAR's Computational and Information Systems Laboratory, sponsored by the National Science Foundation. We are grateful to J.~Hohl for editing the manuscript.
\end{acknowledgements}

\section*{Data Availability}
The data that support the findings of this study are available from the corresponding author upon reasonable request.

\bibliographystyle{elsarticle-harv}
\bibliography{full,addon}

\end{document}